\documentclass[draft,tightenlines,nofootinbib,preprint,aps,eqsecnum,amsmath,amssymb]{revtex4}

\newcommand{\beq}{\begin{equation}}
\newcommand{\eeq}{\end{equation}}
\newcommand{\bea}{\begin{eqnarray}}
\newcommand{\eea}{\end{eqnarray}}

\newcommand{\sgn}{\epsilon}

\begin{document}

\title{Post-Minkowskian Gravity: Dark Matter as a Relativistic Inertial Effect?}

\medskip

\author{Luca Lusanna}

\affiliation{ Sezione INFN di Firenze and ACES Topical Team of ESA\\ Polo Scientifico\\ Via Sansone 1\\
50019 Sesto Fiorentino (FI), Italy\\ Phone: 0039-055-4572334\\
FAX: 0039-055-4572364\\ E-mail: lusanna@fi.infn.it}

\begin{abstract}

Talk at the {\it 1st Mediterranean Conference in Classical and
Quantum Gravity}, held in the Orthodox Academy of Crete in Kolymbari
(Greece) from Monday, September 14th to Friday, September 18th,
2009.

\bigskip

A review is given of the theory of non-inertial frames (with the
associated inertial effects and  the study of the non-relativistic
limit) in Minkowski space-time, of parametrized Minkowski theories
and of the rest-frame instant form of dynamics for isolated systems
admitting a Lagrangian description. The relevance and gauge
equivalence of the clock synchronization conventions for the
identification of the instantaneous 3-spaces (Euclidean only in
inertial frames) are described.

Then this formalism is applied to tetrad gravity in globally
hyperbolic, asymptotically Minkowskian space-times without
super-translations, where the equivalence principle implies the
absence of global inertial frames. The recently discovered York
canonical basis, diagonalizing the York-Lichnerowicz approach,
allows to identify the gauge variables (inertial effects in general
relativity) and the tidal ones (the gravitational waves of the
linearized theory) and to clarify the meaning of the Hamilton
equations. The role of the gauge variable ${}^3K$, the trace of the
extrinsic curvature of the non-Euclidean 3-space (the York time not
existing in Newton theory), as a source of inertial effects is
emphasized. After the presentation of preliminary results on the
linearization of tetrad gravity in the family of non-harmonic
3-orthogonal gauges with a free value of ${}^3K$, we define
post-Minkowskian gravitational waves (without post-Newtonian
approximations on the matter sources) propagating in a non-Euclidean
3-space, emphasizing the non-graviton-like aspects of gravity. It is
conjectured that dark matter may be explained as a relativistic
inertial effect induced by ${}^3K$: it would simulate the need to
choose a privileged gauge connected with the observational
conventions for the description of matter.

\end{abstract}

\maketitle

As shown in Refs. \cite{1A,2A} the recent developments in atom
interferometry and in space physics around the Earth require not
only a consistent special relativistic treatment of atomic physics
in inertial frames in Minkowski space-time and its extension to
non-inertial frames, but also the inclusion of the weak field limit
of gravity at least in the framework of Einstein's general
relativity applied to asymptotically Minkowskian space-times. Once
this description is under control, we must quantize the matter both
in inertial and non-inertial frames in Minkowski space-time. In this
way it is possible to have light rays propagating along flat null
geodesics (then replaced by general relativistic null geodesics in
the weak field limit), avoiding to have "photons" reduced to states
with two polarizations in a 2-dimensional Hilbert space without a
carrier as it is done in the non-relativistic theory of entanglement
and in the associated experiments. The existing inclusion of
electro-magnetism at the order $1/c$ made by atomic physics destroys
the Galilei group and does not allow a consistent definition of the
Poincare' one, namely a consistent special relativistic formulation
of atomic physics in Minkowski space-time. \medskip

While in Galilei space-time both Newtonian time and the Euclidean
3-space, with the associated notion of spatial distance, are {\it
absolute} and Maxwell equations do not exist, in Minkowski
space-time only the space-time is {\it absolute} and there is no
intrinsic definition of an instantaneous 3-space where to formulate
the Cauchy problem for such equations (the fields cannot be put into
a box, because no relativistically covariant box exists). The only
intrinsic notion of special relativity is the fixed Minkowski
light-cone, describing the locus of incoming and outgoing radiation.
The situation becomes more complex in general relativity, where we
have a deformed light-cone varying from a point to the other of
space-time due to gravity. As a consequence, the pre-requisite  to
put control on the predictability for future times (a well posed
Cauchy problem for the relevant partial differential equations) is
the solution of the problem of {\it clock synchronization}, i.e. of
the convention needed to introduce a notion of simultaneity
identifying an instantaneous 3-space. Moreover the solution must be
such that the transition from a simultaneity convention to another
one has to be formulated as a gauge transformation (so that physical
results are not influenced by the convention, which only modifies
the appearances of phenomena) and it must be possible to extend it
to the general relativistic framework, where also the space-time
becomes dynamical and the equivalence principle implies the
existence (at best) of only global non-inertial frames (the inertial
ones exist only locally near an observer in free fall).

\medskip

Regarding the problem of clock synchronization in presence of
gravity near the Earth let us underline the relevance of the ACES
mission of ESA \cite{3A}, programmed for 2013. It will make possible
a measurement of the gravitational redshift of the Earth from the
two-way link among a microwave clock (PHARAO) on the Space Station
and similar clocks on the ground: the proposed microwave link should
make possible the control of effects on the scale of 5 picoseconds.
This will be a test of post-Newtonian gravity in the framework of
Einstein's geometrical view of gravitation: the redshift is a
measure of the $1/c^2$ deviation of post-Newtonian null geodesics
from Minkowski ones. This is a non-perturbative effect (requiring a
re-summation of the whole perturbative expansion) for every quantum
field theory, which has to fix the background (and therefore the
associated light-cones) to be able to define the quantum Fock space
of the theory.\medskip

See Ref.\cite{4A} for the effects of the general relativistic
description of gravity upon atom interferometry.
\medskip

As already said, in special relativity there is no notion of
simultaneity, of instantaneous 3-spaces and of spatial distance. The
light postulates say that the two-way (or round-trip; only one clock
is involved) velocity of light is a) isotropic and b) constant (a
standard constant $c$ replaces the standard of length in existing
relativistic metrology). The one-way velocity of light between two
observers depends on how their clocks are synchronized (in general
is not isotropic and point-dependent). Usually one uses Einstein's
convention for clock synchronization: an inertial observer A send a
ray of light at $x^o_i$ towards the (in general accelerated)
observer B; the ray is reflected towards A at a point P of B
world-line and then reabsorbed by A at $x^o_f$; by convention P is
synchronous with the mid-point between emission and absorption on
A's world-line, i.e. $x^o_P = x^o_i + {1\over 2}\, (x^o_f - x^o_i)$.
This convention selects the Euclidean instantaneous 3-spaces $x^o =
ct = const.$ of the inertial frames centered on A. Only in this case
the one-way velocity of light between A and B coincides with the
two-way one, $c$. However, if the observer A is accelerated, the
convention breaks down, because if {\it only} the world-line of the
accelerated observer A (the {\it 1+3 point of view}) is given, then
the only way for defining instantaneous 3-spaces is to identify them
with the Euclidean tangent planes orthogonal to the 4-velocity of
the observer (the local rest frames). But these planes (they are
tangent spaces not 3-spaces!) will intersect each other at a
distance from A's world-line of the order of the acceleration
lengths of A, so that all the either linearly or rotationally
accelerated frames, centered on accelerated observers, based either
on Fermi coordinates or on rotating ones, will develop {\it
coordinate singularities}. Therefore their approximated notion of
instantaneous 3-spaces cannot be used for a well-posed Cauchy
problem for Maxwell equations. See Refs\cite{5A,6A,7A} for more
details and a rich bibliography on these topics. \medskip

{\it Parametrized Minkowski theories} \cite{9A}, \cite{6A},
\cite{5A}, with the associated {\it inertial and non-inertial
rest-frame instant form of dynamics}, solve these problems and,
together with the results of Refs.\cite{10A,11A,12A}, allow to get a
formulation of relativistic atomic physics \cite{13A}, \cite{14A},
\cite{7A}, both in inertial and non-inertial frames of Minkowski
space-time.

\medskip

To formulate this theory without the coordinate singularities of the
1+3 point of view, we need the {\it 3+1 point of view} \cite{6A}, in
which we assign: a) the world-line of an arbitrary time-like
observer; b) an admissible 3+1 splitting of Minkowski space-time,
namely a nice foliation with space-like instantaneous 3-spaces (i.e.
a clock synchronization convention). This allows to define a {\it
global non-inertial frame} centered on the observer and to use
observer-dependent Lorentz-scalar {\it radar 4-coordinates}
$\sigma^A = (\tau ;\sigma^r)$, where $\tau$ is a monotonically
increasing function of the proper time of the observer and
$\sigma^r$ are curvilinear 3-coordinates on the 3-space
$\Sigma_{\tau}$ having the observer as origin. If $x^{\mu} \mapsto
\sigma^A(x)$ is the coordinate transformation from the inertial
Cartesian 4-coordinates $x^{\mu}$ to radar coordinates, its inverse
$\sigma^A \mapsto x^{\mu} = z^{\mu}(\tau ,\sigma^r)$ defines the
{\it embedding} functions $z^{\mu}(\tau ,\sigma^r)$ describing the
3-spaces $\Sigma_{\tau}$ as embedded 3-manifold into Minkowski
space-time. The induced 4-metric on $\Sigma_{\tau}$ is the following
functional of the embedding ${}^4g_{AB}(\tau ,\sigma^r) =
[z^{\mu}_A\, \eta_{\mu\nu}\, z^{\nu}_B](\tau ,\sigma^r)$, where
$z^{\mu}_A = \partial\, z^{\mu}/\partial\, \sigma^A$ and
${}^4\eta_{\mu\nu} = \sgn\, (+---)$ is the flat metric ($\sgn = \pm
1$ according to either the particle physics $\sgn = 1$ or the
general relativity $\sgn = - 1$ convention). While the 4-vectors
$z^{\mu}_r(\tau ,\sigma^u)$ are tangent to $\Sigma_{\tau}$, so that
the unit normal $l^{\mu}(\tau ,\sigma^u)$ is proportional to
$\epsilon^{\mu}{}_{\alpha \beta\gamma}\, [z^{\alpha}_1\,
z^{\beta}_2\, z^{\gamma}_3](\tau ,\sigma^u)$, we have
$z^{\mu}_{\tau}(\tau ,\sigma^r) = [N\, l^{\mu} + N^r\,
z^{\mu}_r](\tau ,\sigma^r)$ ($N(\tau ,\sigma^r) = \sgn\,
[z^{\mu}_{\tau}\, l_{\mu}](\tau ,\sigma^r)$ and $N_r(\tau ,\sigma^r)
= - \sgn\, g_{\tau r}(\tau ,\sigma^r)$ are the lapse and shift
functions).\medskip

The foliation is nice and admissible if it satisfies the conditions:
\hfill\break
 1) $N(\tau ,\sigma^r) > 0$ in every point of
$\Sigma_{\tau}$ (the 3-spaces never intersect, avoiding the
coordinate singularity of Fermi coordinates);\hfill\break
 2) $\sgn\, {}^4g_{\tau\tau}(\tau ,\sigma^r) > 0$, so to avoid the
 coordinate singularity of the rotating disk, and with the positive-definite 3-metric
${}^3g_{rs}(\tau ,\sigma^u) = - \sgn\, {}^4g_{rs}(\tau ,\sigma^u)$
having three positive eigenvalues (these are the M$\o$ller
conditions \cite{5A,7A});\hfill\break
 3) all the 3-spaces $\Sigma_{\tau}$ must tend to the same space-like
 hyper-plane at spatial infinity (so that there are always asymptotic inertial
observers to be identified with the fixed stars).\medskip

These conditions imply that global {\it rigid} rotations are
forbidden in relativistic theories \cite{5A}. In Ref.\cite{7A} there
is the expression of the admissible embedding corresponding to a 3+1
splitting of Minkowski space-time with parallel space-like
hyper-planes (not equally spaced due to a linear acceleration)
carrying differentially rotating 3-coordinates without the
coordinate singularity of the rotating disk. It is the first
consistent global non-inertial frame of this type.\medskip

In the 3+1 point of view the 4-metric ${}^4g_{AB}(\tau ,\vec \sigma
)$ on $\Sigma_{\tau}$ has the components $\sgn\, {}^4g_{\tau\tau} =
N^2 - N_r\, N^r$, $- \sgn\, {}^4g_{\tau r} = N_r = {}^3g_{rs}\,
N^s$, ${}^3g_{rs} = - \sgn\, {}^4g_{rs} = \sum_{a=1}^3\,
{}^3e_{(a)r}\, {}^3e_{(a)s} = {\tilde \phi}^{2/3}\, \sum_{a=1}^3\,
e^{2\, \sum_{\bar b =1}^2\, \gamma_{\bar ba}\, R_{\bar b}}\,
V_{ra}(\theta^i)\, V_{sa}(\theta^i)$), where ${}^3e_{(a)r}(\tau
,\sigma^u)$ are cotriads on $\Sigma_{\tau}$, ${\tilde \phi}^2(\tau
,\sigma^r) = det\, {}^3g_{rs}(\tau ,\sigma^r)$ is the 3-volume
element on $\Sigma_{\tau}$, $\lambda_a(\tau ,\sigma^r) = [{\tilde
\phi}^{1/3}\, e^{\sum_{\bar b =1}^2\, \gamma_{\bar ba}\, R_{\bar
b}}](\tau ,\sigma^r)$ are the positive eigenvalues of the 3-metric
($\gamma_{\bar aa}$ are suitable numerical constants) and
$V(\theta^i(\tau ,\sigma^r))$ are diagonalizing rotation matrices
depending on three Euler angles.\medskip

The components ${}^4g_{AB}$ or the quantities $N$, $N_r$, $\gamma$,
$R_{\bar a}$, $\theta^i$, play the role of the {\it inertial
potentials} generating the relativistic apparent forces in the
non-inertial frame. It can be shown \cite{7A} that the Newtonian
inertial potentials are hidden in the functions $N$, $N_r$ and
$\theta^i$. The extrinsic curvature ${}^3K_{rs}(\tau, \sigma^u) =
[{1\over {2\, N}}\, (N_{r|s} + N_{s|r} - \partial_{\tau}\,
{}^3g_{rs})](\tau, \sigma^u)$, describing the {\it shape} of the
instantaneous 3-spaces of the non-inertial frame as embedded
3-manifolds of Minkowski space-time, is a functional of the
independent inertial potentials ${}^4g_{AB}$.

\medskip

In parametrized Minkowski theories one considers any isolated system
(particles, strings, fields, fluids) admitting a Lagrangian
description, because it allows, through the coupling to an external
gravitational field, the determination of the matter energy-momentum
tensor and of the ten conserved Poincare' generators $P^{\mu}$ and
$J^{\mu\nu}$ (assumed finite) of every configuration of the system.
Then one replaces the external gravitational 4-metric in the coupled
Lagrangian with the 4-metric $g_{AB}(\tau ,\sigma^r)$ of an
admissible 3+1 splitting of Minkowski space-time and  the matter
fields with new ones knowing the instantaneous 3-spaces
$\Sigma_{\tau}$. For instance a Klein-Gordon field $\tilde \phi (x)$
will be replaced with $\phi(\tau ,\sigma^r) = \tilde \phi (z(\tau
,\sigma^r))$; the same for every other field. Instead for a
relativistic particle with world-line $x^{\mu}(\tau )$ we must make
a choice of its energy sign: then it will be described by
3-coordinates $\eta^r(\tau )$ defined by the intersection of the
world-line with $\Sigma_{\tau}$: $x^{\mu}(\tau ) = z^{\mu}(\tau
,\eta^r(\tau ))$. Differently from all the previous approaches to
relativistic mechanics, the dynamical configuration variables are
the 3-coordinates $\eta^r_i(\tau)$ and not the world-lines
$x^{\mu}_i(\tau)$ (to rebuild them in an arbitrary frame we need the
embedding defining that frame!).\medskip

With this procedure we get a Lagrangian depending on the given
matter and on the embedding $z^{\mu}(\tau ,\sigma^r)$, which is
invariant under frame-preserving diffeomorphisms. As a consequence,
there are four first-class constraints (an analogue of the
super-Hamiltonian and super-momentum constraints of canonical
gravity) implying that the embeddings $z^{\mu}(\tau ,\sigma^r)$ are
{\it gauge variables}, so that all the admissible non-inertial or
inertial frames are gauge equivalent, namely physics does {\it not}
depend on the clock synchronization convention and on the choice of
the 3-coordinates $\sigma^r$: only the appearances of phenomena
change by changing the notion of instantaneous 3-space.

\medskip

As already said, in general relativity the space-time is no more
absolute: it becomes dynamical and it is described by ten fields,
i.e. by the 4-metric tensor ${}^4g_{\mu\nu}(x)$. To get its
Hamiltonian description we will use the same 3+1 point of view and
the radar 4-coordinates employed in special relativity. But now the
admissible embeddings $x^{\mu} = z^{\mu}(\tau, \sigma^r)$ are not
dynamical variables: instead their gradients $z^{\mu}_A(\tau,
\sigma^r)$ give the transition coefficient from radar to world
4-coordinates, ${}^4g_{AB}(\tau, \sigma^r) = [z^{\mu}_A\,
z^{\nu}_B](\tau, \sigma^r)\, {}^4g_{\mu\nu}(z(\tau, \sigma^r))$. As
shown in Ref. \cite{8A}, the dynamical nature of space-time implies
that each solution of Einstein's equations dynamically selects a
preferred 3+1 splitting of the space-time, namely in general
relativity the instantaneous 3-spaces (and therefore the associated
clock synchronization convention) are dynamically determined.

\medskip

If we restrict ourselves to inertial frames, we can define the {\it
inertial rest-frame instant form of dynamics for isolated systems}
by choosing the 3+1 splitting corresponding to the intrinsic
inertial rest frame of the isolated system centered on an inertial
observer: the instantaneous 3-spaces, named Wigner 3-space due to
the fact that the 3-vectors inside them are Wigner spin-1 3-vectors
\cite{6A,9A}, are orthogonal to the conserved 4-momentum $P^{\mu}$
of the configuration. In Ref.\cite{7A} there is the extension to
admissible non-inertial rest frames, where $P^{\mu}$ is orthogonal
to the asymptotic space-like hyper-planes to which the instantaneous
3-spaces tend at spatial infinity. This non-inertial family of 3+1
splittings is the only one admitted by the asymptotically
Minkowskian space-times covered by canonical gravity formulation
discussed below. \medskip

In the inertial rest frames we can get the explicit form of the
Poincare' generators (in particular of the Lorentz boosts, which,
differently from the Galilei ones, are interaction dependent).  We
can also give the final solution to the old problem of the
relativistic extension of the Newtonian center of mass of an
isolated system. In its rest frame there are {\it only} three
notions of collective variables, which can be built by using {\it
only} the Poincare' generators (they are {\it non-local} quantities
knowing the whole $\Sigma_{\tau}$) \cite{10A}: the canonical
non-covariant Newton-Wigner center of mass (or center of spin), the
non-canonical covariant Fokker-Pryce center of inertia and the
non-canonical non-covariant M$\o$ller center of energy. All of them
tend to the Newtonian center of mass in the non-relativistic limit.
See Ref.\cite{6A} for the M$\o$ller non-covariance world-tube around
the Fokker-Pryce 4-vector identified by these collective variables.
As shown in Refs.\cite{10A,11A,12A} these three variables can be
expressed as known functions of the rest time $\tau$, of the
canonically conjugate Jacobi data (frozen Cauchy data) $\vec z =
Mc\, {\vec x}_{NW}(0)$ (${\vec x}_{NW}(\tau )$ is the standard
Newton-Wigner 3-position) and $\vec h = \vec P/Mc$, of the invariant
mass $Mc = \sqrt{\sgn\, P^2}$ of the system and of its rest spin
${\vec {\bar S}}$. It is convenient to center the inertial rest
frame on the Fokker-Pryce inertial observer. As a consequence, every
isolated system (i.e. a closed universe) can be visualized as a
decoupled non-covariant collective (non-local) pseudo-particle
described by the frozen Jacobi data $\vec z$, $\vec h$ carrying a
{\it pole-dipole structure}, namely the invariant mass and the rest
spin of the system, and with an associated {\it external}
realization of the Poincare' group. The universal breaking of
Lorentz covariance is connected to this decoupled non-local
collective variable and is irrelevant because all the dynamics of
the isolated system leaves inside the Wigner 3-spaces and is
Wigner-covariant. In each Wigner 3-space $\Sigma_{\tau}$ there is a
{\it unfaithful internal} realization of the Poincare' algebra,
whose generators are built by using the energy-momentum tensor of
the isolated system. While the internal energy and angular momentum
are $Mc$ and ${\vec {\bar S}}$ respectively, the internal 3-momentum
vanishes: it is the rest frame condition. Also the internal Lorentz
boost (whose expression in presence of interactions is given for the
first time) vanishes: this condition identifies the covariant
non-canonical Fokker-Pryce center of inertia as the natural inertial
observer origin of the 3-coordinates $\sigma^r$ in each Wigner
3-space. As a consequence \cite{13A} there are three pairs of second
class (interaction-dependent) constraints eliminating the internal
3-center of mass and its conjugate momentum inside the Wigner
3-spaces \cite{14A}: this avoids a double counting of the collective
variables and allows to re-express the dynamics only in terms of
internal Wigner-covariant relative variables. As a consequence, we
find that disregarding the unobservable center of mass all the
dynamics is described only by relative variables: this is a form of
{\it weak relationism} without the heavy foundational problem of
approaches like the one in Ref.\cite{15A}.
\medskip

In the case of relativistic particles the reconstruction of their
world-lines requires a complex interaction-dependent procedure
delineated in Ref.\cite{12A}. See Ref.\cite{13A} for the comparison
with the other formulations of relativistic mechanics developed for
the study of the problem of {\it relativistic bound states} and
Ref.\cite{7A} for the extension to non-inertial frames, especially
to the rest-frame ones.

\medskip

In Ref.\cite{1A} there is a review of the formulation of
relativistic atomic physics with the electro-magnetic field in the
radiation gauge (so that Coulomb potential governs the mutual
interaction of the particles; in Ref.\cite{7A} there is its
expression in non-inertial frames) given in Refs.\cite{11A,13A,14A}.
The main result is that, given a suitable regularization of
self-energies by means of the use of Grassmann-valued electric
charges, there exist a canonical transformation leading to particles
mutually interacting with the sum of Coulomb and Darwin potentials
plus a {\it decoupled} transverse radiation field in the rest frame.
Therefore at the classical level for the first time we get the
identification of the Darwin potential (till now obtainable only
from the instantaneous approximations to the Bethe-Salpeter equation
in the theory of relativistic bound states) and a way out from the
Haag theorem.\medskip

Moreover in Ref.\cite{16A} we are able to give a consistent
quantization of relativistic mechanics in absence of the
electro-magnetic field in the inertial rest frame. In it we quantize
the frozen Jacobi data of the canonical non-covariant decoupled
center of mass and the Wigner-covariant relative variables on the
Wigner hyper-plane. In this way we avoid the causality problems of
the Hegerfeldt theorem \cite{17A} (the instantaneous spreading of
wave packets). Since the center of mass is decoupled, its
non-covariance is irrelevant: like for the wave function of the
universe, who will observe it?\medskip

Due to the need of clock synchronization for the definition of the
instantaneous 3-spaces, this Hilbert space $H = H_{com, HJ} \otimes
H_{rel}$ ($H_{com, HJ}$ is the Hilbert space of the external center
of mass in the Hamilton-Jacobi formulation, while $H_{rel}$ is the
Hilbert space of the internal relative variables) is not unitarily
equivalent to $H_1 \otimes H_2 \otimes ...$, where $H_i$ are the
Hilbert spaces of the individual particles. As a consequence, at the
relativistic level the zeroth postulate of non-relativistic quantum
mechanics does not hold: the Hilbert space of composite systems is
not the tensor product of the Hilbert spaces of the sub-systems.
\medskip

This quantization can be extended to the class of global
non-inertial frames with space-like hyper-planes as 3-spaces and
differentially rotating 3-coordinates defined in Ref.\cite{7A}. As
shown in Ref.\cite{18A}, in this {\it multi-temporal quantization}
we quantize only the 3-coordinates $\eta^r_i(\tau)$ of the particles
and {\it not} the inertial effects (like the Coriolis and
centrifugal ones): they remain c-numbers describing the appearances
of phenomena! We get results compatible with atomic spectra.\medskip

Instead the quantization of {\it fields} in non-inertial frames is
an {\it open} problem due to the no-go theorem of Ref.\cite{18A}
showing the existence of obstructions to the unitary evolution of a
massive quantum Klein-Gordon field between two space-like surfaces
of Minkowski space-time. Its solution, i.e. the identification of
all the 3+1 splittings allowing unitary evolution, will be a
prerequisite to any attempt to quantize canonical gravity taking
into account the equivalence principle (global inertial frames do
not exist!). We have already found global non-inertial frames where
the quantization leads to unitary evolution, but we do not yet know
the full class of non-inertial frames evading the no-go theorem.
\medskip

As reviewed in Ref.\cite{1A}, the quantization defined in
Ref.\cite{16A} leads to a first formulation of a theory for {\it
relativistic entanglement}. The non validity of the zeroth postulate
and the {\it non-locality} of Poincare' generators imply a {\it
kinematical non-locality} and a {\it kinematical spatial
non-separability} introduced by special relativity, which reduce the
relevance of {\it quantum non-locality} in the study of the
foundational problems of quantum mechanics which have to be
rephrased in terms of relative variables.

\bigskip
\bigskip

Let us now consider Einstein's general relativity where space-time
is dynamical, gravity is described by the 4-metric tensor and the
equivalence principle says that global inertial frames do not exist.
In it, differently from every other field theory, the 4-metric
tensor has a double role: a) like in electro-magnetism and
Yang-Mills theory it is a (pre)potential for the gravitational
field; b) it also determines the chrono-geometrical structure of
space-time through the line element $ds^2 = {}^4g_{\mu\nu}\,
dx^{\mu}\, dx^{\nu}$. Therefore it teaches relativistic causality to
the other fields. In particular it says to massless particles which
are the allowed trajectories (null geodesics) in each point of
space-time. As already said and shown in Ref.\cite{2A}, the ACES
mission of ESA \cite{3A} will give the first precision measurement
of the gravitational redshift of the geoid, namely of the $1/c^2$
deformation of Minkowski light-cone caused by the geo-potential. In
every quantum field theory, where the definition of the Fock space
requires the use of the fixed light-cone of the background, so that
property b) is lost and gravity is reduced to the spin-2 massless
graviton, this is a non-perturbative effect requiring the
resummation of the perturbative expansion.\medskip

Since all the properties of the standard model of elementary
particles are connected with properties of the representations of
the Poincare' group in inertial frames of Minkowski space-time, we
look for a family of space-times admitting the presence of a
Poincare' algebra. As a consequence, we shall restrict ourselves to
globally hyperbolic, asymptotically Minkowskian at spatial infinity,
topologically trivial space-times, for which a well defined
Hamiltonian formulation of gravity is possible if we replace the
Hilbert action with the ADM one. The 4-metric tends in a suitable
way to the flat Minkowski 4-metric ${}^4\eta_{\mu\nu}$ at spatial
infinity: having an {\it asymptotic} Minkowskian background we can
avoid to split the 4-metric in the bulk in a background plus
perturbations in the weak field limit.\medskip

In developing the Hamiltonian formulation, as already said, we use
the same 3+1 formalism previously introduced for parametrized
Minkowski theories: the basic dynamical variable is now the 4-metric
${}^4g_{AB}(\tau, \sigma^r)$ and not the embedding $z^{\mu}(\tau,
\sigma^r)$. Since tetrad gravity is more natural for the coupling of
gravity to the fermions, the 4-metric is decomposed in terms of
cotetrads, ${}^4g_{AB} = E_A^{(\alpha)}\,
{}^4\eta_{(\alpha)(\beta)}\, E^{(\beta)}_B$ ($(\alpha)$ are flat
indices; the cotetrads $E^{(\alpha)}_A$ are the inverse of the
tetrads $E^A_{(\alpha)}$ connected to the world tetrads by
$E^{\mu}_{(\alpha)}(x) = z^{\mu}_A(\tau, \sigma^r)\,
E^A_{(\alpha)}(z(\tau, \sigma^r))$), and the ADM action, now a
functional of the 16 fields $E^{(\alpha)}_A(\tau, \sigma^r)$, is
taken as the action for ADM tetrad gravity. This leads to an
interpretation of gravity based on a congruence of time-like
observers endowed with orthonormal tetrads: in each point of
space-time the time-like axis is the  unit 4-velocity of the
observer, while the spatial axes are a (gauge) convention for
observer's gyroscopes.
\medskip

The kinematical Poincare' group connecting inertial frames in
special relativity and its enlargement to the group of
frame-preserving diffeomorphisms required for the treatment of
non-inertial frames in parametrized Minkowski theories are now
replaced by the full spatio-temporal diffeonorphism group enlarged
with the O(3,1) gauge group of the Newman-Penrose approach (the
extra gauge freedom acting on the tetrads in the tangent space of
each point of space-time and reducing from 16 to 10 the number of
independent fields like in metric gravity). The relativity principle
of special relativity is replaced with the principle of general
covariance (invariance in form of physical laws). Let us remark that
the problem of pseudo-tensors in the attempt to describe the
gravitational energy, is already present in the non-inertial frames
of Minkowski space-time, since they are connected to inertial frames
by frame-preserving diffeomorphisms.

\medskip

See the papers of Ref.\cite{20A} for this reformulation of canonical
gravity.\medskip

If the direction-independent boundary conditions on the 4-metric and
its conjugate momenta are such to kill {\it super-translations}
\cite{21A}, the SPI group of asymptotic symmetries \cite{22A} is
reduced to the ADM Poincare' group. At the Hamiltonian level a well
posed definition of Poisson brackets and of variational principles
requires the addition of the DeWitt surface term at spatial infinity
to the Dirac Hamiltonian. As shown in Refs.\cite{20A}, with the
previous boundary conditions this term turns out to be the {\it
strong} ADM energy (a flux through a 2-surface at spatial infinity),
which is equal to the {\it weak} ADM energy (expressed as a volume
integral over the 3-space) plus constraints. Therefore in this
family of space-times there is {\it not a frozen picture}, like in
the family of spatially compact without boundary space-times
considered in loop quantum gravity, where the Dirac Hamiltonian is a
combination of constraints.

\medskip

Moreover the absence of super-translations implies that the
non-inertial rest frames are the only family of 3+1 splittings
admitted by these asymptotically Minkowskian space-times, since the
asymptotic Euclidean 3-spaces turn out to be orthogonal to the ADM
4-momentum. Therefore the instantaneous 3-spaces  are non-inertial
rest frames of the 3-universe and admit asymptotic inertial
observers (to be identified with the fixed stars of star
catalogues). If $\epsilon^{\mu}_A$ are a set of asymptotic flat
tetrads, the simplest embedding adapted to the 3+1 splitting of
space-time is $x^{\mu} = z^{\mu}(\tau, \sigma^r) = x^{\mu}(\tau) +
\epsilon^{\mu}_r\, \sigma^r$; if the time-like observer origin of
the spatial radar coordinates is the inertial, with respect to the
asymptotic 4-metric, observer $x^{\mu}(\tau) = x_o^{\mu} +
\epsilon^{\mu}_A\, \sigma^A$, then we have ${}^4g_{AB}(\tau,
\sigma^r) = \epsilon^{\mu}_A\, \epsilon^{\nu}_B\,
{}^4g_{\mu\nu}(x)$. As a consequence, the 3-universe contained in an
instantaneous 3-space can be described as a decoupled non-covariant
non-observable pseudo-particle carrying a pole-dipole structure,
whose mass and spin are those identifying the configuration of the
"gravitational field plus matter" isolated system present in the
3-universe. As a consequence, the ADM Poincare' algebra with weak
generators ${\check P}^A_{ADM}$, ${\check J}^{AB}_{ADM}$ expressed
in radar 4-coordinates has to be considered as the unfaithful {\it
internal} Poincare' algebra of special relativity: a) ${\check
P}^r_{ADM} \approx 0$ are the rest-frame conditions; b) ${\check
J}^{\tau r}_{ADM} \approx 0$ are the gauge-fixings eliminating the
internal center of mass inside the 3-space. The weak ADM energy and
angular momentum define the rest mass and spin of the 3-universe. In
absence of matter Christodoulou - Klainermann space-times \cite{23A}
are compatible with this description.\medskip

With this kind of formalism we can get a deparametrization of
general relativity: if we switch off the Newton constant and we
choose the flat Minkowski 4-metric in Cartesian coordinates as
solution of Einstein's equations, we get the description of the
matter present in the 3-universe in the inertial rest frames of
Minkowski space-time  with the weak ADM Poincare' group collapsing
in the Poincare' group of particle physics.\medskip

\medskip

This framework was developed in the works in Refs.\cite{20A}. The
cotetrads were connected to cotetrads adapted to the 3+1 splitting
of space-time (so that the time-like tetrad is boosted to coincide
with the unit normal to the instantaneous 3-space $\Sigma_{\tau}$,
as it is done in Schwinger time gauges) by standard Wigner boosts
for time-like vectors of parameters $\varphi_{(a)}(\tau, \sigma^r)$,
$a=1,2,3$: $E_A^{\alpha)} = L^{(\alpha)}{}_{(\beta)}(
\varphi_{(a)})\, {\buildrel o\over E}_A^{(\beta)}$. The adapted
cotetrads have the following expression in terms of cotriads
${}^3e_{(a)r}$ on $\Sigma_{\tau}$ and of  the lapse $N = 1 + n$ and
shift $n_{(a)} = N^r\, {}^3e_{(a)r}$ functions: ${\buildrel o\over
E}_{\tau}^{(o)} = 1 + n$, ${\buildrel o\over E}_r^{(o)} = 0$,
${\buildrel o\over E}_{\tau}^{(a)} = n_{(a)}$, ${\buildrel o\over
E}_r^{(a)} = {}^3e_{(a)r}$. The 4-metric becomes ${}^4g_{\tau\tau} =
\sgn\, [(1 + n)^2 - \sum_a\, n^2_{(a)}]$, ${}^4g_{\tau r} = - \sgn\,
\sum_a\, n_{(a)}\, {}^3e_{(a)r}$, ${}^4g_{rs} = - \sgn\, {}^3g_{rs}
= - \sgn\, \sum_a\, {}^3e_{(a)r}\, {}^3e_{(a)s}$. The 16
configurational variables in the ADM action are $\varphi_{(a)}$, $1
+ n$, $n_{(a)}$, ${}^3e_{(a)r}$. There are ten primary constraints
(the vanishing of the 7 momenta of boosts, lapse and shift variables
plus three constraints describing the rotation on the flat indices
$(a)$ of the cotriads) and four secondary ones (the
super-Hamiltonian and super-momentum constraints): all of them are
first class in the phase space spanned by 16+16 fields. This implies
that there are 14 gauge variables describing {\it inertial effects}
and 2 canonical pairs of physical degrees of freedom describing the
{\it tidal effects} of the gravitational field (namely gravitational
waves in the weak field limit). In this canonical basis only the
momenta ${}^3\pi^r_{(a)}$ conjugated to the cotriads are not
vanishing.
\medskip

Then in Ref.\cite{24A} we have found a canonical transformation to a
canonical basis adapted to ten of the first class constraints. It
implementes the York map of Ref.\cite{25A} and diagonalizes the
York-Lichnerowicz approach \cite{26A}. Its final form is
($\alpha_{(a)}(\tau, \sigma^r)$ are angles)

\bea
 &&\begin{minipage}[t]{4 cm}
\begin{tabular}{|ll|ll|l|l|l|} \hline
$\varphi_{(a)}$ & $\alpha_{(a)}$ & $n$ & ${\bar n}_{(a)}$ &
$\theta^r$ & $\tilde \phi$ & $R_{\bar a}$\\ \hline
$\pi_{\varphi_{(a)}} \approx0$ &
 $\pi^{(\alpha)}_{(a)} \approx 0$ & $\pi_n \approx 0$ & $\pi_{{\bar n}_{(a)}} \approx 0$
& $\pi^{(\theta )}_r$ & $\pi_{\tilde \phi} = {{c^3}\over {12\pi G}}\, {}^3K$ & $\Pi_{\bar a}$ \\
\hline
\end{tabular}
\end{minipage}\nonumber \\
 &&{}\nonumber \\
 &&{}\nonumber \\
 &&{}^3e_{(a)r} = R_{(a)(b)}(\alpha_{(c)})\, {}^3{\bar e}_{(b)r} =
 R_{(a)(b)}(\alpha_{(c)})\, V_{rb}(\theta^i)\,
 {\tilde \phi}^{1/3}\, e^{\sum_{\bar a}^{1,2}\, \gamma_{\bar aa}\, R_{\bar a}},\nonumber \\
 &&{}^4g_{\tau\tau} = \sgn\, [(1 + n)^2 - \sum_a\, {\bar n}^2_{(a)}],
 \qquad {}^4g_{\tau r} = - \sgn\, {\bar
 n}_{(a)}\, {}^3{\bar e}_{(a)r},\nonumber \\
 &&{}^4g_{rs} = - \sgn\, {}^3g_{rs} = - \sgn\, {\tilde \phi}^{2/3}\,
 \sum_a\, V_{ra}(\theta^i)\, V_{sa}(\theta^i)\,
 e^{2\, \sum_{\bar a}^{1,2}\, \gamma_{\bar aa}\, R_{\bar
 a}},\nonumber \\
 &&{}\nonumber \\
 \eea

In this York canonical basis the {\it inertial effects} are
described by the arbitrary gauge variables $\alpha_{(a)}$,
$\varphi_{(a)}$, $1 + n$, ${\bar n}_{(a)}$, $\theta^i$, ${}^3K$,
while the {\it tidal effects}, i.e. the physical degrees of freedom
of the gravitational field, by the two canonical pairs $R_{\bar a}$,
$\Pi_{\bar a}$, $\bar a =1,2$. The momenta $\pi_r^{(\theta)}$ and
the 3-volume element $\tilde \phi = \sqrt{det\, {}^3g_{rs}}$ have to
be found as solutions of the super-momentum and super-hamiltonian
(i.e. the Lichmerowicz equation) constraints, respectively.\medskip

The gauge variables $\alpha_{(a)}$, $\varphi_{(a)}$ parametrize the
extra O(3,1) gauge freedom of the tetrads (the gauge freedom for
each observer to choose three gyroscopes as spatial axes and to
choose the law for their transport along the world-line). The gauge
angles $\theta^i$ (i.e. the director cosines of the tangents to the
three coordinate lines in each point of $\Sigma_{\tau}$) describe
the freedom in the choice of the 3-coordinates $\sigma^r$ on each
3-space: their fixation implies the determination of the shift gauge
variables ${\bar n}_{(a)}$, namely the appearances of
gravito-magnetism in the chosen 3-coordinate system.\medskip

The final basic gauge variable is a momentum, namely the trace
${}^3K(\tau ,\sigma^r)$ of the extrinsic curvature (also named the
{\it York time}) of the non-Euclidean 3-space $\Sigma_{\tau}$. While
in Yang-Mills theory all the gauge variables are configuration ones,
the Lorentz signature of space-time implies that ${}^3K$ is a
momentum variable: it is a time coordinate, while $\theta^i$ are
spatial coordinates. While in special relativity the extrinsic
curvature of the 3-space of non-inertial frames is a derived gauge
quantity (the basic inertial potentials are ${}^4g_{AB}[z]$), here
it is an independent gauge variable describing the shape of 3-space
as an embedded 3-manifold in the space-time. This is what remains of
the special relativistic gauge freedom in the choice of the clock
synchronization convention: the other five components of
${}^3K_{rs}$ are determined by $R_{\bar a}$, $\Pi_{\bar a}$,
$\pi_r^{(\theta)}$, $\tilde \phi$ and $\theta^i$. This gauge
variable has no Newtonian counterpart (the Euclidean 3-space is
absolute), because its fixation determines the final shape of the
non-Euclidean 3-space and then the lapse gauge variable (i.e. the
proper time in each point of 3-space). Moreover this gauge variable
gives rise to a negative kinetic term in the weak ADM energy,
vanishing only in the gauges ${}^3K(\tau, \vec \sigma) = 0$.

\medskip

In the York canonical basis the Hamilton equations generated by the
Dirac Hamiltonian $H_D = {\hat E}_{ADM} + (constraints)$ are divided
in four groups: A) the contracted Bianchi identities, namely the
evolution equations for $\tilde \phi$ and $\pi_i^{(\theta)}$ (they
say that given a solution of the constraints on a Cauchy surface, it
remains a solution also at later times); B) the evolution equation
for the four basic gauge variables $\theta^i$ and ${}^3K$: these
equations determine the lapse and the shift functions once the basic
gauge variables are fixed; C) the evolution equations for the tidal
variables $R_{\bar a}$, $\Pi_{\bar a}$; D) the Hamilton equations
for matter, when present.

\medskip

Once a gauge is completely fixed, the Hamilton equations become
deterministic. Given a solution of the super-momentum and
super-Hamiltonian constraints and the Cauchy data for the tidal
variables on an initial 3-space, we can find a solution of
Einstein's equations in radar 4-coordinates adapted to a time-like
observer. To it there is associated a special 3+1 splitting of
space-time with dynamically selected instantaneous 3-spaces in
accord with Ref.\cite{8A}. Then we can get pass to adapted world
4-coordinates ($x^{\mu} = x^{\mu}_o + \epsilon^{\mu}_A\, \sigma^A$)
and we can describe the solution in every 4-coordinate system by
means of 4-diffeomorphisms.

\medskip

In Ref.\cite{27A} we study the coupling of N charged scalar
particles plus the electro-magnetic field to ADM tetrad gravity  in
this class of asymptotically Minkowskian space-times without
super-translations. To regularize the self-energies both the
electric charge and the sign of the energy of the particles are
Grassmann-valued. The introduction of the non-covariant radiation
gauge allows to reformulate the theory in terms of transverse
electro-magnetic fields and to extract the generalization of the
Coulomb interaction among the particles in the Riemannian
instantaneous 3-spaces of global non-inertial frames.

\medskip

After the reformulation of the whole system in the York canonical
basis, we give the restriction of the Hamilton equations and of the
constraints to the family of {\it non-harmonic 3-orthogonal}
Schwinger time gauges, in which the instantaneous Riemannian
3-spaces have a non-fixed trace ${}^3K$ of the extrinsic curvature
but a diagonal 3-metric. This family of gauges is determined by the
gauge fixings $\theta^i(\tau, \sigma^r) \approx 0$ and ${}^3K(\tau,
\sigma^r) \approx (arbitrary\, numerical\, function)$.

\medskip

Starting from the results obtained in Ref.\cite{27A} for this family
of non-harmonic 3-orthogonal Schwinger gauges, it is  possible to
define a consistent {\it linearization} of ADM canonical tetrad
gravity plus matter in the weak field approximation \cite{28A}, to
obtain a formulation of Hamiltonian Post-Minkowskian gravity with
non-flat Riemannian 3-spaces and asymptotic Minkowski background.
This means that the 4-metric tends to the asymptotic Minkowski
metric at spatial infinity, ${}^4g_{AB}\, \rightarrow
{}^4\eta_{AB}$. The decomposition ${}^4g_{AB} = {}^4\eta_{AB} +
{}^4h_{(1)AB}$, with a first order perturbation ${}^4h_{(1)AB}$
vanishing at spatial infinity, is only used for calculations, but
has no intrinsic meaning. Moreover, due to the presence of a
ultra-violet cutoff for matter, we can avoid to make Post-Newtonian
expansions, namely we get fully relativistic expressions. We have
found solutions for the first order quantities
$\pi^{(\theta)}_{(1)r}$, $\tilde \phi = 1 + 6\, \phi_{(1)}$, $1 +
n_{(1)}$, ${\bar n}_{(1)(a)}$. Then we can show that the tidal
variables $R_{\bar a}$ satisfy a wave equation $\Box\, R_{\bar a} =
(known\, functional\, of\, matter)$ with the D'Alambertian
associated to the asymptotic Minkowski 4-metric. Therefore, by using
a no-incoming radiation condition based on the asymptotic Minkowski
light-cone, we get a description of gravitational waves in these
non-harmonic gauges, which can be connected to generalized
TT(transverse traceless) gauges, as retarded functions of the
matter. These gravitational waves do not propagate in inertial
frames of the background (like it happens in the standard harmonic
gauge description), but in non-Euclidean instantaneous 3-spaces
differing from Euclidean 3-spaces at the first order (their
intrinsic 3-curvature is detemined by the gravitational waves) and
dynamically determined by the linearized solution of Einstein
equations. These 3-spaces have a first order extrinsic curvature
(with ${}^3K_{(1)}(\tau, \sigma^r)$ describing the clock
synchronization convention) and a first order modification of
Minkowski light-cone.
\medskip

We can write explicitly the linearized Hamilton equations for the
particles and for the electro-magnetic field: among the forces there
are both the inertial potentials and the gravitational waves. If we
disregard electro-magnetism, we can study the non-relativistic limit
of the particle equations. The preliminary result \cite{28A} is that
the particle 3-coordinates $\eta^r_i(\tau = ct) = {\tilde
\eta}_i^r(t)$ satisfy the equation (${\vec F}_{Newton}$ is the
Newton gravitational force) $m\, {{d^2 {\tilde \eta}^r_i(t)}\over
{dt^2}} = \sum_{j \not= i}\, F^r_{Newton}({\vec {\tilde \eta}}_i(t)
- {\vec {\tilde \eta}}_j(t)) + {1\over c}\, {{d {\tilde
\eta}^r_i(t)}\over {dt}}\, \Big({1\over {\triangle}}\, c^2\,
\partial^2_{\tau}\, {}^3K_{(1)}(\tau = ct, \vec \sigma)\Big)
{|}_{\vec \sigma = {\vec {\tilde \eta}}_i(t)}$. Therefore the
(arbitrary in these gauges) double rate of change in time of the
trace of the extrinsic curvature creates a post-Newtonian damping
(or anti-damping since the sign of ${}^3K_{(1)}$ is not fixed)
effect on the motion of particles. This is a inertial effect not
existing in Newton theory where the Euclidean 3-space is absolute.
\medskip

As a consequence there is the possibility of describing part (or
maybe all) dark matter as a {\it relativistic inertial effect}
determined by the gauge variable ${}^3K(\tau, \sigma^r)$: the
rotation curves of galaxies would then experimentally determine a
preferred choice of the instantaneous 3-spaces by using the freedom
in ${}^3K_{(1)}$ to fit them. This option would differ both from the
non-relativistic MOND approach (where one modifies Newton equations)
and from modified gravity theories like the $f(R)$ ones (where one
gets a modification of Newton potential) and is under investigation.
\medskip

Since the implication of this approach to the dark matter problem is
the existence of privileged gauges, let us add some remarks on the
observables and the gauge problem in general relativity. Since no
one is able to evaluate the Dirac observables of the gravitational
field and of matter in general relativity (see Pons' talk for the
status of the art), let us look at what are the measurable
quantities used for observations when gravity is present.
Gravitational waves are searched by existing, like LIGO and VIRGO on
the Earth surface, or planned, LISA in space, detectors. To study
these detectors one must introduce conventional 4-coordinate systems
and the transformations rules among them: the terrestrial one ITRS
(IERS2003) on the Earth surface, the geodetic celestial one GCRS
(IAU2000) near the Earth and the barycentric celestial one BCRS
(IAU2000) for the Solar System (see the bibliography of
Ref.\cite{2A}). While the barycentric frame is considered as an
approximate inertial Minkowski frame, the geocentric one is obtained
from it by means of a rotation-free Lorentz boost plus
post-Newtonian corrections. Therefore the post-Newtonian description
of matter extended systems in the Solar System is intrinsically
coordinate dependent since it requires the identification of the
trajectory of the object. Satellites (think to GPS) are described
with NASA coordinates first in ITRS ad then in GCRS. The planets in
the Solar System are described in BCRS. The description of stars and
galaxies done by astronomers with their star catalogues are based on
an extension of BCRS, i.e. on a celestial frame which is an inertial
frame of Minkowski space-time reduced to an inertial frame of
Galilei space-time. This is due to the fact that the reconstruction
of a 4-dimensional space-time from the 2-dimensional observations
(light and angles) uses as input the standard cosmological model
(the homogeneous and isotropic FRW solution of Einstein's equations)
with the constant intrinsic 3-curvature put equal to zero, so that
the 3-space is an Euclidean 3-space. As a consequence all the theory
of galactic dynamics is strictly non-relativistic. However this set
of conventions and assumptions lead to the open problems of dark
matter and dark energy and the validity of the standard cosmological
model is under investigation. Since the existence of dark matter is
deduced from non-relativistic Newton theory in Euclidean 3-space,
the conjecture is that we must choose a gauge in general relativity
(where the gauge freedom is the choice of 4-coordinates) consistent
with the observational conventions. Since all the observational
frames have diagonal 3-metric, our family of 3-orthogonal gauges is
reasonable. The conjecture coming from our approach is that all the
data on what is called dark matter should be used to try to find
which value of the York time ${}^3K$ explains the observations if we
accept a notion of non-Euclidean 3-spaces. In this way a privileged
gauge connected to observations would begin to emerge, to be used to
describe the classical physics after the recombination surface.
\medskip

More information on the role played by the York time will derive
from how the following quantities depend upon it (at least in the
linearized theory): a) the local proper time of a time-like
observer; b) the redshift of rays of light along null geodesics; c)
the luminosity distance based on the geodesic deviation equation
along null geodesics. The future investigation of these quantities
will also give information on dark energy.

\medskip

Finally, if we will replace the matter with a perfect fluid (for
instance dust), this  will alow us to try to see whether the York
canonical basis can help in developing the back-reaction approach
\cite{bb} to dark energy, according to which dark energy is a
byproduct of the non-linearities of general relativity when one
considers spatial mean values on large scales to get a cosmological
description of the universe taking into account the inhomogeneity of
the observed universe. Since in the 3-orthogonal gauges all the
relevant quantities are 3-scalars, it is possible to study the mean
value of all the Hamilton equations and not only of the two
considered by Buchert and to evaluate them explicitly in the
linearized formulation. This will be done in the previous privileged
gauge with the recombination surface as Cauchy surface.

\vfill\eject

\vfill\eject


\begin{thebibliography}{}


\bibitem{1A}L.Lusanna, {\it Towards Relativistic Atomic Physics and
Post-Minkowskian Gravitational Waves}, talk at the Workshop on {\it
Gravitational Waves Detection with Atom Interferometry}, held in
Firenze at the Galileo Galilei Institute for Theoretical Physics,
February 23-24, 2009 (0908.0209).

\bibitem{2A}L.Lusanna, {\it The Chrono-Geometrical Structure of
General Relativity and Clock Synchronization}, talk at the {\it
First Colloquium Scientific and Fundamental Aspects of the Galileo
Programme}, Toulouse, October 1-4, 2007 (0708.0490).

\bibitem{3A} L.Cacciapuoti and C.Salomon, {\it ACES: Mission Concept and
Scientific Objective}, 28/03/2007, ESA document, Estec
(ACES{}\_{}Science{}\_{}v1{}\_{}printout.doc).\hfill\break
 L.Blanchet, C.Salomon, P.Teyssandier and P.Wolf, {\it Relativistic
Theory for Time and Frequency Transfer to Order $1/c^3$},
Astron.Astrophys. {\bf 370}, 320 (2000).\hfill\break
 L.Lusanna, {\it Dynamical Emergence of 3-Space in General
Relativity: Implications for the ACES Mission}, in Proc. of the 42th
Rencontres de Moriond {\it Gravitational Waves and Experimental
Gravity}, La Thuile (Italy), 11-18 March 2007.\hfill\break
 See also  the talks at the {\it SIGRAV Graduate School on Experimental
Gravitation in Space}(Firenze, September 25-27, 2006)
(http://www.fi.infn.it/GGI-grav-space/egs{}\_{}s.html); at the
Workshop {\it Advances in Precision Tests and Experimental
Gravitation in Space} (Firenze, September 28/30, 2006)
(http://www.fi.infn.it/GGI-grav-space/egs{}\_{}w.html); at the
Workshop "Theoretical Aspects of the ACES Mission" (Firenze, April
29-30, 2008) (ftp://cacciapuoti:In73rn0@ftp.estec.esa.int/ ); at the
Workshop on "ACES and Future GNSS-based Earth Observation and
Navigation" (Muenchen, May 26-27, 2008)
(http://www.iapg.bv.tum.de/12735--~aces~programme.html).


\bibitem{4A}S.Dimopoulous, P.W.Graham, J.M.Hogan and M.A.Kasevich, {\it General
 Relativistic Effects in Atom Interferometry} (arXiv:
 0802.4098).\hfill\break
 S.Dimopoulous, P.W.Graham, J.M.Hogan, M.A.Kasevich and S.Rajendran,
 {\it An Atomic Gravitational Wave Interferometric Sensor (AGIA)}
 (arXiv: 0806.2125).


\bibitem{5A}D.Alba and L.Lusanna, {\it Generalized Radar 4-Coordinates
and Equal-Time Cauchy Surfaces for Arbitrary Accelerated Observers}
(2005),  Int.J.Mod.Phys. {\bf D16}, 1149 (2007) (gr-qc/0501090).

\bibitem{6A}L.Lusanna, {\it The Chrono-Geometrical Structure of Special and General
Relativity: A Re-Visitation of Canonical Geometrodynamics}, lectures
at 42nd Karpacz Winter School of Theoretical Physics: Current
Mathematical Topics in Gravitation and Cosmology, Ladek, Poland,
6-11 Feb 2006, Int.J.Geom.Methods in Mod.Phys. {\bf 4}, 79 (2007).
(gr-qc/0604120).\hfill\break
 L.Lusanna, {\it The Chronogeometrical
Structure of Special and General Relativity: towards a
Background-Independent Description of the Gravitational Field and
Elementary Particles} (2004), in {\it General Relativity Research
Trends}, ed. A.Reiner, Horizon in World Physics vol. 249 (Nova
Science, New York, 2005) (gr-qc/0404122).



\bibitem{7A}D.Alba and L.Lusanna, {\it Charged Particles and the
Electro-Magnetic Field in Non-Inertial Frames}, (arXiv 0812.3057),
to appear in Int.J.Geom.Methods in Physics in the form of the
following two papers {\it Charged Particles and the Electro-Magnetic
Field in Non-Inertial Frames: I.  Admissible 3+1 Splittings of
Minkowski Spacetime and the Non-Inertial Rest Frames (0908.0213) and
II. Applications: Rotating Frames, Sagnac Effect, Faraday Rotation,
Wrap-up Effect (0908.0215)}.




\bibitem{8A} L.Lusanna and M.Pauri,
{\it Explaining Leibniz equivalence as difference of non-inertial
Appearances: Dis-solution of the Hole Argument and physical
individuation of point-events}, History and Philosophy of Modern
Physics {\bf 37}, 692 (2006) (gr-qc/0604087); {\it The Physical Role
of Gravitational and Gauge Degrees of Freedom in General Relativity.
I: Dynamical Synchronization and Generalized Inertial Effects; II:
Dirac versus Bergmann Observables and the Objectivity of
Space-Time}, Gen.Rel.Grav. {\bf 38}, 187 and 229 (2006)
(gr-qc/0403081 and 0407007); {\it Dynamical Emergence of
Instantaneous 3-Spaces in a Class of Models of General Relativity},
to appear in the book {\it Relativity and the Dimensionality of the
World}, ed. A. van der Merwe (Springer Series Fundamental Theories
of Physics) (gr-qc/0611045).



\bibitem{9A}L. Lusanna, {\it The N- and 1-Time Classical Descriptions of N-Body
Relativistic Kinematics and the Electromagnetic Interaction}, Int.
J. Mod. Phys. {\bf A12}, 645 (1997).

\bibitem{10A} D.Alba, L.Lusanna and M.Pauri, \textit{New Directions in
Non-Relativistic and Relativistic Rotational and Multipole
Kinematics for N-Body and Continuous Systems} (2005), in
\textit{Atomic and Molecular Clusters: New Research}, ed.Y.L.Ping
(Nova Science, New York, 2006) (hep-th/0505005).\hfill\break
  D.Alba, L.Lusanna and M.Pauri, \textit{Centers of Mass and Rotational
Kinematics for the Relativistic N-Body Problem in the Rest-Frame
Instant Form}, J.Math.Phys. \textbf{43}, 1677-1727 (2002)
(hep-th/0102087).\hfill\break
  D.Alba, L.Lusanna and M.Pauri,
\textit{ Multipolar Expansions for Closed and Open Systems of
Relativistic Particles} , J. Math.Phys. \textbf{46}, 062505, 1-36
(2004) (hep-th/0402181).


\bibitem{11A} H.Crater and L.Lusanna, {\it The Rest-Frame Darwin Potential from
 the Lienard-Wiechert Solution in the Radiation Gauge},
 Ann.Phys.(N.Y.) {\bf 289}, 87 (2001)(hep-th/0001046).\hfill\break
 D.Alba, H.Crater and L.Lusanna, {\it The Semiclassical
 Relativistic Darwin Potential for Spinning Particles in the
 Rest-Frame Instant Form: Two-Body Bound States with Spin 1/2
 Constituents}, Int.J.Mod.Phys. {\bf A16}, 3365 (2001) (hep-th/0103109).


\bibitem{12A} D.Alba, H.W.Crater and L.Lusanna, \textit{Hamiltonian
Relativistic Two-Body Problem: Center of Mass and Orbit
Reconstruction}, J.Phys. {\bf A40}, 9585 (2007) (gr-qc/0610200).


\bibitem{13A}D.Alba, H.W.Crater and L.Lusanna, {\it Towards Relativistic
Atom Physics. I. The Rest-Frame Instant Form of Dynamics and a
Canonical Transformation for a system of Charged Particles plus the
Electro-Magnetic Field}, to appear in Canad.J.Phys. (arXiv:
0806.2383).



\bibitem{14A}D.Alba, H.W.Crater and L.Lusanna, {\it Towards Relativistic
Atom Physics. II. Collective and Relative  Relativistic Variables
for a System of Charged Particles plus the Electro-Magnetic Field},
to appear in Canad.J.Phys.(0811.0715).

\bibitem{15A}C.Rovelli, {\it Relational Quantum Mechanics},
Int.J.Theor.Phys. {\bf 35}, 1637 (1996) (quant-ph/9609002).
\hfill\break
 C.Rovelli and M.Smerlak, {\it Relational EPR}, Found.Phys. {\bf 37},
 427 (2007)(quant-ph/0604064).


\bibitem{16A}D.Alba, H.W.Crater and L.Lusanna, {\it Relativistic
Quantum Mechanics and Relativistic Entanglement in the Rest-Frame
Instant Form of Dynamics}, (arXiv 0907.1816).

\bibitem{17A}G.C.Hegerfeldt, {\it Remark on Causality and Particle
Localization}, Phys.Rev. {\bf D10}, 3320 (1974); {\it Instantaneous
Spreading and Einstein Causality in Quantum Theory}, Ann.Phys.Lpz.
{\bf 7}, 716 (1998) (quant-ph/9809030); {\it Causality, Particle
Localization and Positivity of the Energy},  in {\it Irreversibility
and Causality in Quantum Theory - Semigroups and Rigged Hilbert
Spaces}, eds. A.Bohm, H.D.Doebner and P.Kielanowski, Lecture Notes
in Physics 504, p.238 (Springer , NewYork, 1998) (quant-ph/9806036).


\bibitem{18A}D.Alba and L.Lusanna, {\it Quantum Mechanics in
Non-Inertial Frames with a Multi-Temporal Quantization Scheme: I)
Relativistic Particles}, Int.J.Mod.Phys. {\bf A21}, 2781 (2006)
(hep-th/0502060)\hfill\break
 D.Alba, {\it Quantum Mechanics in Non-Inertial Frames with a
Multi-Temporal Quantization Scheme: II) Non-Relativistic Particles},
Int.J.Mod.Phys. {\bf A21}, 3917 (2006) (hep-th/0504060).




\bibitem{19A}Torre, C.G. and Varadarajan, M. {\it Functional Evolution of
Free Quantum Fields}, Clas. Quantum Grav. {\bf 16}, 2651-2668
(1999).\hfill\break
 A.Arageorgis, J.Earman and L.Ruetsche, {\it Weyling the Time
 Away: the Non-Unitary Implementability of Quantum Field Dynamics
 o n Curved Spacetimes}, Studies in History and
 Philosophy of Modern Physics {\bf 33}, 151 (2002).

\bibitem{20A}L.Lusanna, {\it The Rest-Frame Instant Form of Metric Gravity},
Gen.Rel.Grav. {\bf 33}, 1579 (2001)(gr-qc/0101048).\hfill\break
 L.Lusanna and S.Russo, {\it A New Parametrization for Tetrad Gravity},
Gen.Rel.Grav. {\bf 34}, 189 (2002)(gr-qc/0102074).\hfill\break
R.DePietri, L.Lusanna, L.Martucci and S.Russo, {\it Dirac's
Observables for the Rest-Frame Instant Form of Tetrad Gravity in a
Completely Fixed 3-Orthogonal Gauge}, Gen.Rel.Grav. {\bf 34}, 877
(2002) (gr-qc/0105084).\hfill\break
 J.Agresti, R.De Pietri, L.Lusanna and L.Martucci, {\it
Hamiltonian Linearization of the Rest-Frame Instant Form of Tetrad
Gravity in a Completely Fixed 3-Orthogonal Gauge: a Radiation Gauge
for Background-Independent Gravitational Waves in a Post-Minkowskian
Einstein Spacetime}, Gen.Rel.Grav. {\bf 36}, 1055 (2004)
(gr-qc/0302084).\hfill\break
 J.Agresti, R.De Pietri, L.Lusanna and
L.Martucci, {\it Hamiltonian Linearization of the Rest-Frame Instant
Form of Tetrad Gravity in a Completely Fixed 3-Orthogonal Gauge: a
Radiation Gauge for Background-Independent Gravitational Waves in a
Post-Minkowskian Einstein Spacetime}, Gen.Rel.Grav. {\bf 36}, 1055
(2004) (gr-qc/0302084).

\bibitem{21A}T.Regge and C.Teitelboim, {\it Role of Surface Integrals in the
Hamiltonian Formulation of General Relativity}, Ann.Phys.(N.Y.) {\bf
88}, 286 (1974).
 \hfill\break
 R.Beig and \'O Murchadha, {\it The Poincaré Group as the Symmetry Group of
 Canonical General Relativity}, Ann.Phys.(N.Y.) {\bf 174}, 463
 (1987).\hfill\break
 L.B.Szabados, {\it On the Roots of Poincare' Structure of Asymptotically
 Flat Space-Times}, Class. Quantum Grav. {\bf 20}, 2627 (2003) (gr-qc/0302033).

\bibitem{22A}R.M.Wald, {\it General Relativity} (Chicago Univ. Press, Chicago,
1984).




\bibitem{23A}D.Christodoulou and S.Klainerman, {\it The Global Nonlinear
Stability of the Minkowski Space} (Princeton, Princeton, 1993).






\bibitem{24A}D.Alba and L.Lusanna, {\it The York Map as a Shanmugadhasan
Canonical Transformationn in Tetrad Gravity and the Role of
Non-Inertial Frames in the Geometrical View of the Gravitational
Field}, Gen.Rel.Grav. {\bf 39}, 2149 (2007) (gr-qc/0604086, v2).

\bibitem{25A}J.Isenberg and J.E.Marsden, {\it The York Map is a
Canonical Transformation}, J.Geom.Phys. {\bf 1}, 85 (1984).



\bibitem{26A}I.Ciufolini and J.A.Wheeler, {\it Gravitation and
Inertia} (Princeton Univ.Press, Princeton, 1995).




\bibitem{27A}D.Alba and L.Lusanna, {\it The Einstein-Maxwell-Particle
System in the York Canonical Basis of ADM Tetrad Gravity: I) The
Equations of Motion in Arbitrary Schwinger Time Gauges.}, 2009
(arXiv 0907.4087).

\bibitem{28A}D.Alba and L.Lusanna, {\it The Einstein-Maxwell-Particle
System in the York Canonical Basis of ADM Tetrad Gravity: II)  The
Weak Field Approximation in the 3-Orthogonal Gauges and Hamiltonian
Post-Minkowskian Gravity: the N-Body Problem and Gravitational Waves
with Asymptotic Background.}, in preparation.


\bibitem{bb}T.Buchert, {\it Dark Energy from Structure: a Status Report},
Gen.Rel.Grav. {\bf 40}, 467 (2008) (0707.2153).






\end{thebibliography}
\end{document}